\documentclass[twocolumn]{aastex63}

\usepackage{amsmath,amssymb}
\usepackage{array}
\newcolumntype{C}{>{\centering\arraybackslash}p{5em}}
\usepackage{colortbl}

\newcommand{\DMunit}{$\rm pc\,cm^{-3}$}
\newcommand{\VCSreso}{$10$-kHz/$100$-$\mu$s}
\newcommand{\DataObs}{$80$-$220$\,MHz}
\def\PSR{J2241$-$5236}

\shorttitle{High time resolution study of \PSR}
\shortauthors{Kaur et al.}

\begin{document}
\title[]{\uppercase{A high time resolution study of the millisecond pulsar {\PSR} at frequencies below $300$ MHz}}




\author{D. Kaur} 
\affiliation{International Centre for Radio Astronomy Research (ICRAR), Curtin University, Bentley, WA 6102, Australia}
\affiliation{ARC Centre of Excellence for All-Sky Astrophysics (CAASTRO), Australia}

\author{N. D. R. Bhat} 
\affiliation{International Centre for Radio Astronomy Research (ICRAR), Curtin University, Bentley, WA 6102, Australia}
\affiliation{ARC Centre of Excellence for All-Sky Astrophysics (CAASTRO), Australia}

\author{S. E. Tremblay} 
\affiliation{International Centre for Radio Astronomy Research (ICRAR), Curtin University, Bentley, WA 6102, Australia}
\affiliation{ARC Centre of Excellence for All-Sky Astrophysics (CAASTRO), Australia}

\author{R. M. Shannon}
\affiliation{Centre for Astrophysics and Supercomputing, Swinburne University of Technology, P.O. Box 218, Hawthorn, VIC 3122, Australia}
\affiliation{ARC Centre of Excellence for Gravitational Wave Discovery (OzGrav)}

\author{S. J. McSweeney}
\affiliation{International Centre for Radio Astronomy Research (ICRAR), Curtin University, Bentley, WA 6102, Australia}
\affiliation{ARC Centre of Excellence for All-Sky Astrophysics (CAASTRO), Australia}

\author{S. M. Ord}
\affiliation{CSIRO Astronomy and Space Science, P.O. Box 76, Epping, NSW 1710, Australia}

\author{A. P. Beardsley}
\affiliation{School of Earth and Space Exploration, Arizona State University, Tempe, AZ 85287, USA}

\author{B. Crosse}
\affiliation{International Centre for Radio Astronomy Research (ICRAR), Curtin University, Bentley, WA 6102, Australia}

\author{D. Emrich}
\affiliation{International Centre for Radio Astronomy Research (ICRAR), Curtin University, Bentley, WA 6102, Australia}

\author{T. M. O. Franzen}
\affiliation{International Centre for Radio Astronomy Research (ICRAR), Curtin University, Bentley, WA 6102, Australia}

\author{ L. Horsley}
\affiliation{International Centre for Radio Astronomy Research (ICRAR), Curtin University, Bentley, WA 6102, Australia}

\author{  M. Johnston-Hollitt}
\affiliation{International Centre for Radio Astronomy Research (ICRAR), Curtin University, Bentley, WA 6102, Australia}

\author{   D. L. Kaplan}
\affiliation{Department of Physics, University of Wisconsin--Milwaukee, Milwaukee, WI 53201, USA}

\author{   D. Kenney}
\affiliation{International Centre for Radio Astronomy Research (ICRAR), Curtin University, Bentley, WA 6102, Australia}

\author{    M. F. Morales}
\affiliation{Department of Physics, University of Washington, Seattle, WA 98195, USA}

\author{   D. Pallot}
\affiliation{International Centre for Radio Astronomy Research, University of Western Australia, Crawley 6009, Australia}

\author{    K. Steele}
\affiliation{International Centre for Radio Astronomy Research (ICRAR), Curtin University, Bentley, WA 6102, Australia}

\author{  S. J. Tingay}
\affiliation{International Centre for Radio Astronomy Research (ICRAR), Curtin University, Bentley, WA 6102, Australia}
\affiliation{ARC Centre of Excellence for All-Sky Astrophysics (CAASTRO), Australia}

\author{   C. M. Trott} 
\affiliation{International Centre for Radio Astronomy Research (ICRAR), Curtin University, Bentley, WA 6102, Australia}
\affiliation{ARC Centre of Excellence for All-Sky Astrophysics (CAASTRO), Australia}

\author{    M. Walker} 
\affiliation{International Centre for Radio Astronomy Research (ICRAR), Curtin University, Bentley, WA 6102, Australia}

\author{    R. B. Wayth}
\affiliation{International Centre for Radio Astronomy Research (ICRAR), Curtin University, Bentley, WA 6102, Australia}
\affiliation{ARC Centre of Excellence for All-Sky Astrophysics (CAASTRO), Australia}

\author{   A. Williams}
\affiliation{International Centre for Radio Astronomy Research (ICRAR), Curtin University, Bentley, WA 6102, Australia}

\author{  C. Wu}
\affiliation{International Centre for Radio Astronomy Research, University of Western Australia, Crawley 6009, Australia}

  

\begin{abstract}
One of the major challenges for pulsar timing array (PTA) experiments is the mitigation of the effects of the turbulent interstellar medium (ISM) from timing data. These can potentially lead to measurable delays and/or distortions in the pulse profiles and scale strongly with the inverse of the radio frequency. Low-frequency observations are therefore highly appealing for characterizing them. However, in order to achieve the necessary time resolution to resolve profile features of short-period millisecond pulsars, phase-coherent de-dispersion is essential, especially at frequencies below $300$\,MHz. We present the lowest-frequency ($80$-$220$\,MHz), coherently de-dispersed detections of one of the most promising pulsars for current and future PTAs, PSR {\PSR}, using our new beam-former software for the MWA's voltage capture system (VCS), which reconstructs the time series at a much higher time resolution  of $\sim 1\,\mu$s by re-synthesizing the recorded voltage data at {\VCSreso} native resolutions. Our data reveal a dual-precursor type feature in the pulse profile that is either faint or absent in high-frequency observations from Parkes. The resultant high-fidelity detections have enabled dispersion measure (DM) determinations with very high precision, of the order of $(2$-$6)\times10^{-6}$\,{\DMunit}, owing to the microsecond level timing achievable for this pulsar at the MWA's low frequencies. This underscores the usefulness of low-frequency observations for probing the ISM toward PTA pulsars and informing optimal observing strategies for PTA experiments. 
\end{abstract}


\section{INTRODUCTION} \label{sec:introduction}
Millisecond pulsars (MSPs) are important targets for high-precision pulsar timing array (PTA) experiments, which involve regular monitoring of pulse arrival times from a celestial distribution of pulsars to search for ultra-low frequency (nHz) gravitational waves (e.g., \citealp{2013MancheaterPPTA, 2013Demorest, 2011VanHaasteren, 2015Janssen, 2018BailesPTA}). To achieve the required $\rm \sim 100\,ns$ timing precision, it is important to ascertain various factors contributing to the timing noise budget (e.g., \citealp{2010C&S}). PTA efforts over the past few years suggest that the interstellar medium (ISM) propagation effects on pulsar signals are one of the major contributing factors and may ultimately limit the detection sensitivity of PTAs if they are not accurately measured and corrected for in timing measurements (\citealp{2017Shannon&Cordes, 2017Jones, 2016Levin, 2016Cordes}).

The magnitudes of ISM effects scale strongly with the inverse of observing frequency. Therefore, PTAs generally observe pulsars at higher frequencies ($\sim$\,1-2\,GHz) to reduce the effects of the ISM (\citealp{2018Lam}). Two dominant effects are: $1)$ dispersion, arising from propagation through intervening ionized gas, quantified as the integrated electron density along the line of sight of the pulsar, and is called the dispersion measure (DM); $2)$ pulse broadening, arising from multipath propagation in the turbulent ISM. The dispersion delay scales as $\nu^{-2}$, and pulse broadening time as $\sim$\,$\nu^{-4}$ (e.g., \citealp{2004Bhat}), where $\nu$ is the observing frequency. Low-frequency observations of pulsars can therefore allow a robust characterization of ISM effects, unlike those at higher frequencies where their impact on timing measurements are not easily discernible. Low-frequency observations can also be used to investigate the recently proposed effects such as chromatic (frequency-dependent) DM, arising due to different path lengths traversed, or volumes sampled, at different frequencies, again owing to multipath propagation (\citealp{2016Cordes}). This may also have important implications for timing-array applications, as that would imply that correcting for ISM effects using low-frequency data is not straightforward. Thus, detailed observational investigations over a wide range of frequencies will be very instructive.
 
Developments in low-frequency radio instrumentation and advances in computing over the past decade have revitalized pulsar astronomy at low frequencies. Instruments such as the Murchison Widefield Array (MWA; \citealp{2013Tingay}), the Low Frequency Array (LOFAR; \citealp{2013van}), and the Long Wavelength Array (LWA; \citealp{2012Taylor}) have been built, and operational, and are opening up new opportunities for observing pulsars at frequencies below $\rm 300\,MHz$.
The MWA is also a low frequency ($80$-$300$\,MHz) precursor to the Square Kilometer Array (SKA). It has been recently upgraded, with a doubling of the number of tiles and maximum baseline, i.e. a $256$-tile array extending out to $\sim$\,6\,km (\citealp{2018Wayth}). With its voltage capture system (VCS; \citealp{2015Tremblay}), which records data at {\VCSreso} resolutions, the MWA has been producing a wide variety of pulsar science (e.g., \citealp{2017Sammy, 2017Meyer, 2017Mengyao}), including characterization of the ISM towards important southern-sky PTA targets (e.g., \citealp{2014Bhat, 2016Bhat, 2018Bhat}).

For MSPs with short rotation periods (P\,$\lesssim$\,5\,ms; where P is the period) and narrow pulse profiles, the $10$-kHz channelization of VCS limits their detection, especially at the lower frequencies of the MWA. For example, a pulsar with a DM of $10$\,{\DMunit} will be smeared to $1.62$\,ms at $80$\,MHz. Fundamentally, this hampers the ability to resolve fine temporal features in the pulse profiles of MSPs. Thus, we require the ability for higher time resolution and phase-coherent de-dispersion to remove temporal smearing at the low frequencies of the MWA.

In this paper, we present early science results from our new and improved beamformer software, which allows re-synthesis of $1.28$-MHz/$0.78$-$\mu$s voltage samples from {\VCSreso} native resolutions of the VCS, by performing an inversion of the polyphase filterbank (PFB) operation prior to VCS recording. This is demonstrated using our observations of PSR {\PSR}, an important PTA target. The pulsar was discovered in targeted searches toward unassociated Fermi-LAT candidates, and has a period of $\rm 2.18\,ms$ and a DM of $11.41$\,{\DMunit} (\citealp{2011Keith}). The new beamformer has enabled the pulsar detections down to $\sim$\,80\,MHz, thereby allowing us to study the ISM and spectral properties of the pulsar.
In $\mathsection\,2$ we describe the details of our observations. In $\mathsection\,3$ and $\mathsection\,4$ we summarize data processing and analysis, and in $\mathsection\,5$ we present our results. Our conclusions are summarized in $\mathsection\,6$.

\section{OBSERVATIONS}\label{sec:observations}
Observations were made at three epochs, each separated by nearly $\sim$\,1 year, with the MWA-VCS, which records $24$ coarse channels, each of which has been finely channelized to $10$\,kHz using a $12$-tap PFB. It records both polarizations from a maximum of $128$ tiles, and has an aggregate data rate of $\sim$\,28\,TB\,$\rm hr^{-1}$. The $\rm 30.72\,MHz$ observing bandwidth currently feasible with the MWA can flexibly be recorded in either a contiguous band or can be split into $\rm 24\times1.28\,MHz$ coarse channels, which can be placed anywhere across the nominal operating frequency range ($80$-$300$\,MHz). This distributed channel mode is particularly suitable for exploring frequency-dependent effects such as pulse profile evolution, scintillation and scattering (\citealp{2018Bhat, 2019Kirsten}). The VCS data are transported to the Pawsey supercomputing facility, where all further processing, including calibration and beamforming, are carried out. Further details of the observations are summarized in Table~\ref{tab:MWA_observations}.

\begin{deluxetable}{ccccc}
\tablecaption{Observational details of PSR {\PSR} \label{tab:MWA_observations}}
\tablehead{ \colhead{MJD} & \colhead{Frequency}   & \colhead{Channel} & \colhead{Observing} & \colhead{$\rm N_{tile}$} \tablenotemark{a} \\
	    \colhead{}    & \colhead{range (MHz)} & \colhead{setup}   & \colhead{duration (s)}  }
\startdata
    57288 & 170-200 & 1$\times$30.72 & 4800 & 114\\
    57707 & 80-220  & 12$\times$2.56 \tablenotemark{b} & 1800 & 126\\
    57979 & 80-220  & 12$\times$2.56 \tablenotemark{b} & 3600 & 121\\
\enddata
\tablenotetext{a}{No. of tiles used while producing calibration solutions.}
\tablenotetext{b}{Two consecutive $1.28$\,MHz channels joined together.}
\end{deluxetable}

\section{DATA PROCESSING}\label{sec:data processing}
The VCS data can be processed to form a coherent beam by applying suitable phase rotations to each voltage stream and then summing them to generate a phased array signal (\citealp{2019Ord}). It involves accounting for the polarimetric response for each tile and frequency channel by applying appropriate cable and geometric delays and complex gains (i.e. amplitudes and phases). Data are then calibrated using standard calibrators (e.g., Hydra\,A and $\rm 3C444$), recorded in pointed observations prior to pulsar observations. The calibration solutions are generated for each tile and frequency channel using the Real Time System (RTS; \citealp{2008Mitchell}), and tiles that produced poor calibration solutions are flagged. The full processing pipeline is implemented on the Galaxy cluster at the Pawsey Supercomputing Centre. These data are then processed using the standard pulsar data analysis packages {\tt DSPSR} (\citealp{2011Van&Bailes}) and {\tt PSRCHIVE} (\citealp{2004Hotan, 2012Van}) to generate multi-channel pulse profiles for further analysis.

\subsection{Re-constructing the high-time resolution time series}\label{sec:time series}

High-quality MSP detections require coherent de-dispersion, in which the received voltage signal is deconvolved using the transfer function of the ISM (\citealp{1975Hankins}). While this technique has been routinely implemented in pulsar instrumentation around the world (e.g., \citealp{2011Van&Bailes}), it was not possible earlier with the MWA due to our fine $10$-kHz channelization and the filterbank format of beamformed data (writen in full-Stokes PSRFITS). With the new beamformer software, which reconstructs the data at a much higher time resolution by re-synthesizing the coarse-channel voltage samples (by effectively inverting the fine PFB operation that produced the fine channels), it has become possible to generate data that are amenable for coherent de-dispersion.

The MWA fine channelization is implemented via a PFB with a $12$-tap Kaiser window function, which is one of a class of window functions that allows theoretically perfect inversion of the PFB channelization process (\citealp{1983Cro, 1995Opp})\footnote{In practise, integer quantization in the analysis and/or synthesis process means the reconstruction will not be perfect.}.
The synthesis filterbank is achieved by first up-sampling the output of each $10$-kHz fine channel to the initial time resolution (a factor of $128$) and then passing them through a synthesis filter, which we choose to be the same filter used in the forward PFB stage, with an appropriately chosen phase ramp applied to each fine channel. Then each channel is low-pass filtered to remove the $127$ images of the spectrum created by the up-sampling. The re-synthesized coarse channels have a time resolution of $\sim$\,0.78\,$\rm \mu s$ and can be coherently de-dispersed to produce high-quality detections of MSPs at the MWA's frequencies. Further details of this are described in an upcoming publication (McSweeney et al. in prep).

\section{ANALYSIS}\label{sec:analysis}
\subsection{DM determination}\label{sec:DM determination}
Figure \ref{fig:mwa_full_band} presents MWA detections of PSR {\PSR} across the {\DataObs} range. As evident from the figure, the pulsar is visibly bright across the $\sim100$-$200$\,MHz frequency range. Using the best detection (signal-to-noise ratio; S/N\,$\sim70$), we created an analytic template, which is a noise-free profile model, composed of von Mises functions, generated using the {\tt PAAS} utility of {\tt PSRCHIVE}. 
The analytic template is then cross-correlated with the observed profiles to obtain the times of arrival (ToAs) using the {\tt PSRCHIVE} package.
The ToAs are then analyzed using the pulsar timing package $\tt TEMPO2$ (\citealp{2006Hobbs}) to determine the DM. We adopted the latest available timing solution of the pulsar and only fit for DM in this analysis. As our observing system is configured as 24 individual frequency sub-bands, we derived the DM using a set 24 ToAs for every single observation.

As seen in Figure \ref{fig:mwa_full_band}, the pulse width is relatively unchanged throughout the observed frequency range\footnote{The measured $10\%$ pulse widths $\sim$\,150\,$\rm \mu s$ across our {\DataObs} range.}, and therefore for the purpose of the DM analysis, we assume that there is no significant pulse profile evolution within our observing band. Ideally, the profile evolution needs to be considered (cf.\,\citealp{2019Pennucci}), however, our data quality is not adequate for a detailed analysis. The data are subsequently re-processed using the refined DM to obtain better quality pulse profiles. The related analysis is illustrated in Figure \ref{fig:DM_before_after}, which shows pulsar signal across {\DataObs}, with 12$\times$2.56\,MHz channels joined together. This detection is from MJD 57979 and corresponds to the brightest among all our observations. These profiles at 80-220\,MHz are then averaged in frequency to generate a standard profile centered at $\rm 150\,MHz$, as shown in Figure \ref{fig:mwa_pks_bands}, which is then used as a final template for all observations to obtain the ToAs and DM measurements.

\subsection{Flux density calibration}\label{sec:Flux calibration}

The flux density calibration for our data is carried out using the approach described by \cite{2017Meyer}. This involves estimating the system gain (G) by simulating the tied-array beam pattern and integrating the response across the visible sky for the phased-array configuration used for observing. The tied-array beam pattern is then combined with the Global Sky Model of \citet{2008De} to calculate the antenna temperature ($\rm T_{ant}$). Assuming a radiation efficiency ($\rm \eta$) of unity, the system temperature ($\rm T_{sys}$) is then $\rm T_{sys}\,=\,T_{ant}\,+ T_{rec}$, where $\rm T_{rec}$ is the measured receiver temperature. This process is conducted separately for each sub$\textbf{-}$band, as all of these quantities are strongly frequency dependent. While doing this, we also account for a ``coherency factor" (cf. \citealp{2017Meyer}), for any loss in sensitivity in the coherent beamforming process. 
This information is then used to estimate the flux density for all the observed frequency bands using the radiometer equation (\citealp{2004Lorimer}) and also by factoring in the duty cycle of the pulsar, which is equivalent pulse width (W) divided by the pulse period (W/P), where W is measured as integrated flux over the ``on" pulse divided by the peak amplitude of the pulse.

\section{Results and Discussion}\label{sec:results_discussion}
\begin{figure}[htp]
\epsscale{1.2}
	\plotone{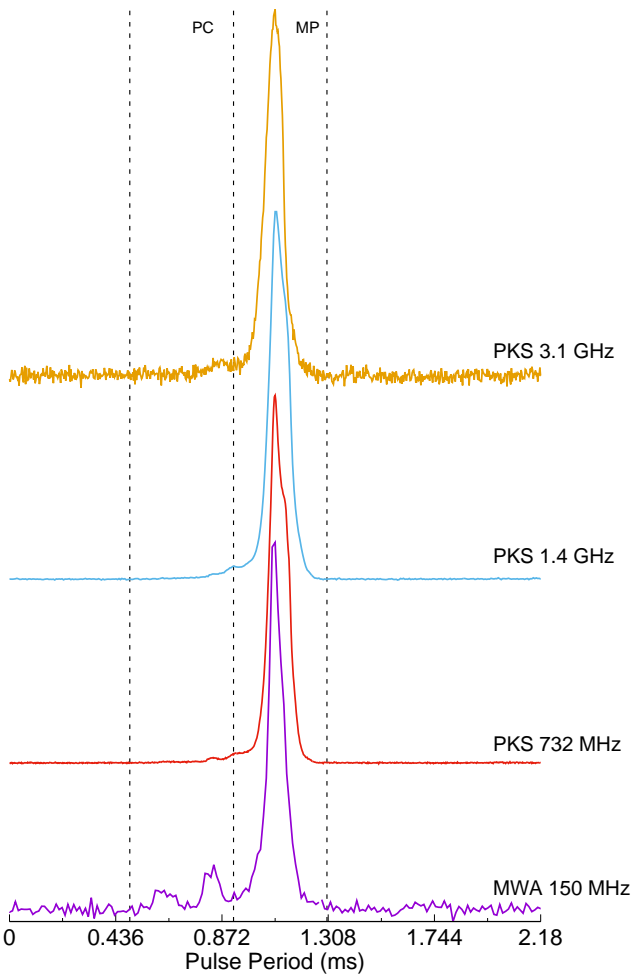}	
	\caption{Integrated pulse profiles of PSR {\PSR} at frequencies from $150$\,MHz to $3.1$\,GHz. The MWA profile has a time resolution of $\sim$\,8.5\,$\rm \mu s$, whereas Parkes profiles (from \citealp{2015Dai}) have a time resolution of $\sim$\,2\,$\rm \mu s$.}
	\label{fig:mwa_pks_bands}
\end{figure}

PSR {\PSR} has so far been primarily studied at frequencies above $\rm 700\,MHz$, with the Parkes telescope (\citealp{2015Dai, 2011Keith}).
Aside from a low S/N detection that was reported by \citet{2017Mengyao}, there are no other low-frequency observation of this pulsar in the published literature. Unlike other bright southern MSPs such as PSRs J0437$-$4715 and J2145$-$0750 (e.g., \citealp{2018Bhat}), PSR {\PSR} has a narrow pulse profile and shows little profile evolution within the MWA's band. Figure \ref{fig:mwa_pks_bands} presents the frequency-averaged profile centered at $\rm 150\,MHz$, and we measure a $10\%$ pulse width ($\rm W_{10}$) of $\sim$\,150\,$\rm \mu s$ at $158$\,MHz, which is comparable to $\sim$130\,$\rm \mu s$ measured at Parkes frequencies ($\sim$\,1-3\,GHz).

\begin{figure}
	\epsscale{1.2}
	\plotone{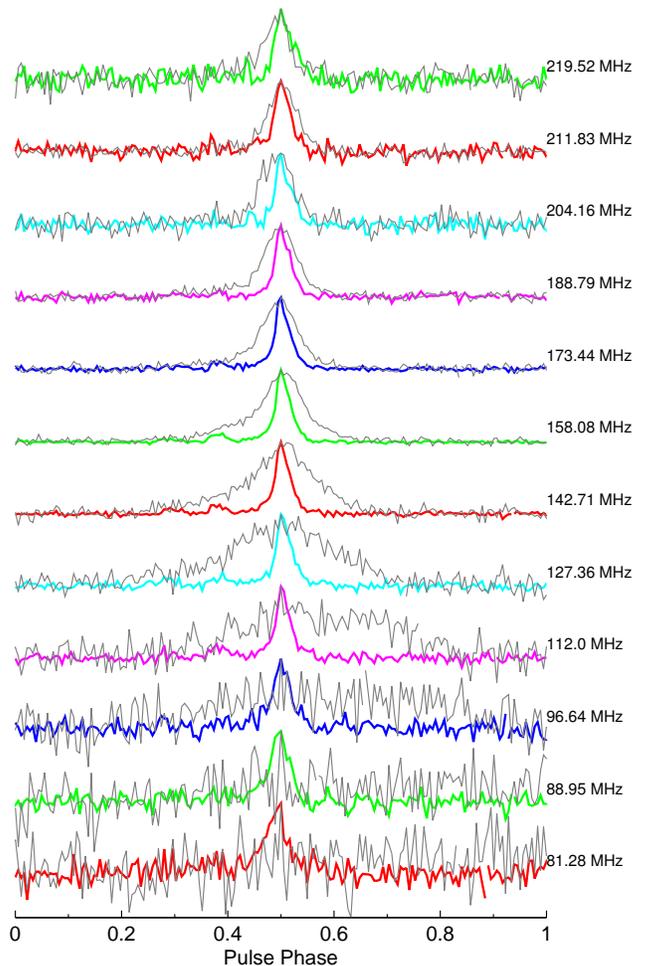}
	\caption{MWA detections of PSR {\PSR} from observations made at MJD $57979$. Data were coherently de-dispersed at each of the 12$\times$2.56\,MHz bands. The pulsar is detected throughout the {\DataObs} range; the resultant profiles (in colour) are overlaid with the detections using the previous system, at $10$\,kHz resolution (in grey), where the
	pulse profile is significantly broadened due to residual dispersive smearing within $10$\,kHz channels ($4\%$ to $\sim$\,80$\%$ of the pulse period).}
	\label{fig:mwa_full_band}
\end{figure}

\begin{figure*}
\centering
\epsscale{1.2}
\plotone{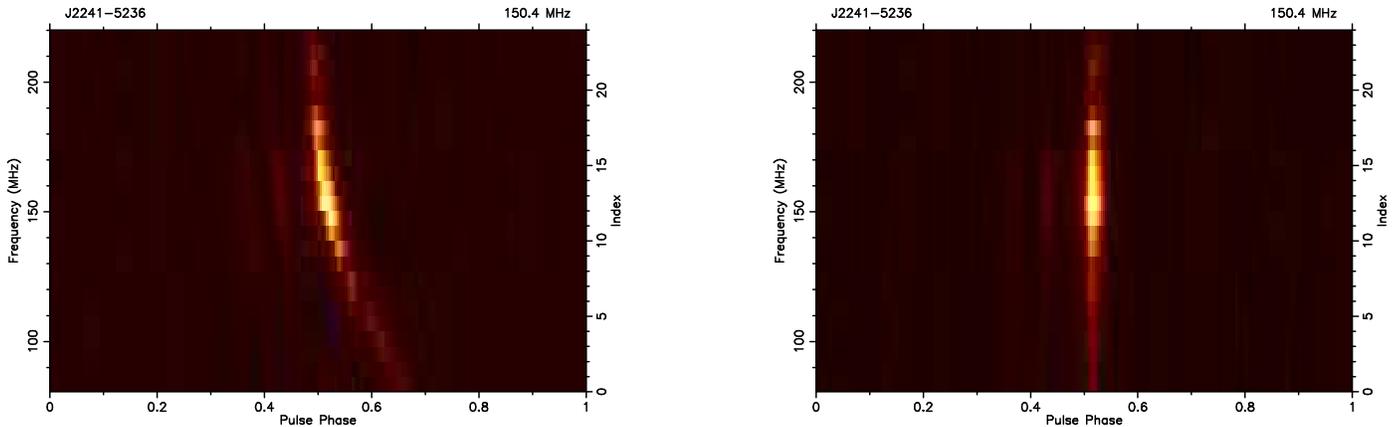}
	\caption{Detection of PSR {\PSR} across the {\DataObs} frequency range of the MWA from observations made at MJD $=57979$. Plots showing pulse intensity as a function of frequency vs. pulse phase, where $12$ sub-bands, each $2.56$\,MHz wide, at separations $\sim 8$-$16$\,MHz, are joined together. Left: data processed with the catalog DM of $11.41085$\,{\DMunit} and the quadratic sweep of $0.4$\,ms implies an excess DM of $0.000667$\,{\DMunit}. Right: data reprocessed with the refined DM of $11.41151$\,{\DMunit}, demonstrating the MWA's sensitivity to subtle DM variations.}
	\label{fig:DM_before_after}
\end{figure*}

In general, MSPs tend to show quite complex and rapid profile evolution with frequency (e.g., \citealp{2015Dai, 2018Bhat}). In the case of J2241$-$5236, while the main pulse shows little evolution with frequency, MWA observations reveal an interesting dual precursor type feature, which is less prominent at Parkes frequencies (see Fig \ref{fig:mwa_pks_bands}). On a closer examination of Figures \ref{fig:mwa_full_band} and \ref{fig:DM_before_after}, it is evident that the precursor emission is persistent over a large fraction of the observed MWA band, where it is nominally seen at $10$-$15\%$ of the pulse peak. However, its relative phase appears to be slightly offset (leftward) compared to that seen at Parkes frequencies. This could possibly be due to the spectral evolution of the profile. Observations at intermediate frequencies ($\sim 300$-$700$\,MHz) would be useful to further investigate this.

\subsection{DM measurements and precision}\label{sec:DM}

We initially de-dispersed the data using the catalogue DM of $11.41085$\,{\DMunit}. This revealed a clear quadratic sweep of approximately $0.4$\,ms as seen in Figure \ref{fig:DM_before_after}. We subsequently reprocessed the data using our refined DM measurements, as summarized in Table\,\ref{tab:MWA_DM}. Our measured DMs are consistently larger than the catalogue value. The average DM from all three observations is $11.411505$\,{\DMunit}, which may suggest an \textbf{``}excess\textbf{"} DM of $0.000655$\,{\DMunit}, compared to the catalogue value as reported in \citet{2011Keith}\footnote{Based on timing data ($\boldsymbol{\sim1}$-2\,GHz) taken over a $\boldsymbol{\sim}$\,15-month time span in 2009-2010.}.

Such subtle changes can in principle arise due to effects such as chromatic (frequency-dependent) DMs, as recently theorised by \citet{2016Cordes}. The chromatic DM being a consequence of multi-path scattering, in general, one may expect a dependence on the degree of scattering. However, our current observations (where we distribute the recording bandwidth to sample multiple spot frequencies across a large frequency range) are not ideal for a meaningful scintillation analysis. In any case, there has been some observational evidence in support of excess DM at low frequencies, e.g. in the case of PSR J2145$-$0750, for which MWA and LWA measurements by \citet{2018Bhat} and \citet{2013Dowell} report a $\delta$DM of 0.006$\pm$0.003\,{\DMunit}. More recently, DM variations have been reported for PSR J2219$+$4754 by \citet{2019Donner} and \citet{2019Lam} from LOFAR observations, and have been attributed to chromatic DMs, although their interpretations differ. With only limited observational efforts to date to investigate this important effect, routine observations with low-frequency instruments such as the MWA and LWA will be especially useful.

PSR {\PSR} is a high-priority target for PTA applications (i.e. along the class of pulsars like J1909$-$3744 and J1744$-$1134). However, to date, there are no published measurements of temporal DM variations (on time scales of $\boldsymbol{\sim}$\,months to years) for this pulsar. As can be seen from Table\,\ref{tab:MWA_DM}, our observations suggest DM variations, $\delta$DM $\sim 10^{-4}$\,$\rm pc\,cm^{-3}$, on timescales of $\sim$\,1-2 years. Since the observed DM variation is subtle, multiple different sources of origin can be considered. For example, the Solar wind is known to produce DM variations up to $\sim$\,$10^{-5}$\,{\DMunit} (e.g., \citealp{2013Ujjwal, 2016Lam}), when observations are made very close to the Sun. However, this seems unlikely, as our observations were made at large Solar elongations ($\gtrsim 80^{\circ}$), and as such the pulsar is far from the ecliptic plane.  There has been some recent evidence that the companion winds can modulate the DM for binary pulsars.  PSR {\PSR} is in a $3.5$-hr, almost circular orbit with a low-mass ($\sim$\,0.01\,$\rm M_{\odot}$) black widow type companion. We may therefore expect subtle variations in DM as a function of the orbital phase, especially in light of the recent work by \citet{2018AnHongjun}, who report orbital modulation in gamma-ray due to intra-binary shock emission. There has been observational evidence for such black-widow like system with an ablating companion producing DM variations, e.g., PSR B1957$+$20 (\citealp{1990Fruchter}). However, in the case of {\PSR}, no eclipses have been seen as the inclination angle is likely large (\citealp{2011Keith}), making this less probable. Moreover, our limited data and a sparse coverage of the orbit refrain us from making definitive comments on this.

The DM precision depends on the frequency lever arm, with the best precision achieved for the longest, which, in our case, is 80-220\,MHz. As can be seen in Table\,\ref{tab:MWA_DM}, this is in the range $(2-6)\times10^{-6}$\,{\DMunit}, depending on the quality of detection. This
is already an order of magnitude better than what is currently achievable at timing frequencies ($3\times10^{-5}$\,{\DMunit}; \citealp{2017Jones}), and even the most recent work of \citet{2019Donner}, who reach a similar level of precision. Since our analysis relied on the use of a single template (and not frequency-dependent templates that are ideal for wide-band timing), there is indeed scope for considerable improvement as we accrue more data with the MWA. Furthermore, the measured DM variation ($\rm \delta DM$), is about $\sim$\,3-10 times larger than our measurement uncertainties (see Table\,\ref{tab:MWA_DM}), highlighting the MWA's high sensitivity to measuring such subtle changes in DMs. 

\begin{deluxetable}{cccc}
\tablecaption{DM measurements from three observed epochs \label{tab:MWA_DM}}
\tablehead{
\colhead{Date of Obs} & \colhead{DM}         & \colhead{$\rm \sigma_{DM}$\tablenotemark{$\dagger$}} & \colhead{Post-fit} \\
\colhead{MJD}         & \colhead({\DMunit})  & \colhead({\DMunit})                                  & \colhead{residual ($\mu$s)}
}
\startdata
 $57288$ & $11.41165(3)$  & ($3$-$9$) $\times$ $10^{-5}$ & $12.6$ \\
 $57707$ & $11.41162(6)$  & ($6$-$9$) $\times$ $10^{-6}$ & $1.4$ \\
 $57979$ & $11.41151(2)$  & ($2$-$7$) $\times$ $10^{-6}$ & $0.8$ \\
\enddata
\tablecomments{
The catalogue DM value for PSR {\PSR} is $11.41085$\,{\DMunit}, with uncertainties of $3\times10^{-5}$\,{\DMunit}  \\
\tablenotemark{$\dagger$} The achievable DM precision ($\sigma _{\rm DM}$) varies depending on the frequency lever arm, which ranges from $\sim$140\,MHz to $\sim$10\,MHz in our analysis (see Table~\ref{tab:MWA_observations} and Figure \ref{fig:mwa_full_band}).
}
\end{deluxetable}

\subsection{Spectral behavior at low frequencies}\label{sec:Spectral behavior at low frequencies}
\begin{figure}[!b]
	\epsscale{1.3}
	\plotone{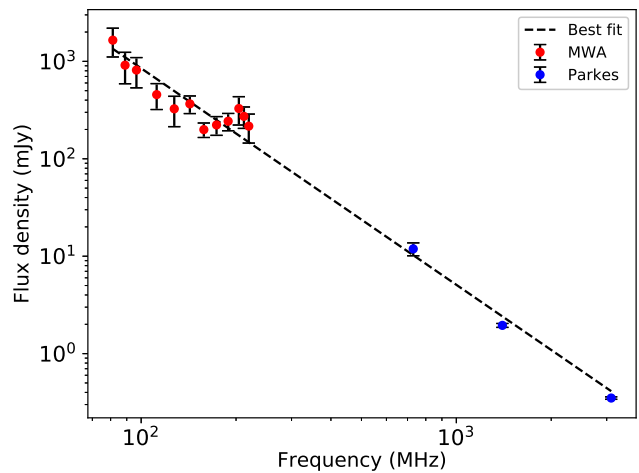}	
	\caption{Flux density measurements of PSR {\PSR} from $\rm 80\,MHz$ to $\rm 3\,GHz$. For the MWA range ({\DataObs}), measurements are averaged over all three observations. The best-fit spectral index is $\rm\alpha_{MP}$\,=\,${-}2.25$\,$\pm$\,$0.10$, compared to $\alpha_{1}$\,=\,${-}2.12$\,$\pm$\,$0.04$ from Parkes measurements.}
	\label{fig:flux_plot}
\end{figure}

The pulsar spectra generally follow a single power\textbf{-}law $\rm S_{\nu}\,\propto\,\nu^{\alpha}$, where $\rm S_{\nu}$ is the flux density at the observing frequency $\nu$ and $\alpha$ is the spectral index. However, there are exceptions, where the spectral form tends to either steepen, or flatten, or even turns over (e.g., \citealp{1998Lorimer, 2000Maron, 2018Jankowski}).
MSP spectra were initially reported to be steeper, compared to those of long-period pulsars (\citealp{1998Toscano}). However, the later work by \citet{2001Kuzmin} claimed a lack of low-frequency spectral turnover for MSPs. More recent work by \citet{2015Dai} suggests a possible turnover for at least a few pulsars, and the work by \citet{2016Kon}, at frequencies below $300$\,MHz, suggests that for many MSPs their measured flux densities are lower than the predicted ones, again hinting at a possible spectral turnover.

Using the procedure explained in $\mathsection\,3.2$ we estimated the flux densities from our integrated pulse profiles. By combining the Parkes and MWA data, we estimated a spectral index for the main pulse (MP) $\rm \alpha_{MP}$\,=\,${-}2.25$\,$\pm$\,0.10. Since the precursor components (PC) are not prominent in the Parkes band, they were excluded from our analysis.
Figure \ref{fig:flux_plot} presents the mean flux densities measured from multiple MWA observations (see Table \ref{tab:MWA_DM}) along with Parkes data from \citet{2015Dai}.
Parkes measurements are from regularly sampled, six-year data set, whereas MWA measurements are from just three observations over a $\sim$\,3\,year time span. Variability in measured flux densities between different observations at the MWA's frequencies is generally within a factor of two, across the entire frequency range, and this is consistent with expectations from refractive scintillation at these low frequencies.\footnote{The modulations due to diffractive scintillation is $\sim$\,30\%, based on our estimated scintillation parameters.} The time scale for refractive interstellar scintillation (RISS), $\tau_{\rm riss}$ = ($\nu / \nu_{\rm diss})\,\tau_{\rm diss}$, where $\nu_{\rm diss}$ and $\tau_{\rm diss}$ are the frequency and time scales for diffractive interstellar scintillation, respectively (\citealp{1990Rickett}). For our observations at 158\,MHz (MJD = 57979), preliminary estimates are $\sim$\,200\,kHz for decorrelation bandwidth ($\nu_{\rm diss}$) and $\sim$\,1000\,s for time scale ($\tau_{\rm diss}$). Thus, we estimate a refractive time scale, $\tau_{\rm riss}$ $\sim$\,9\,days. The small number of observations and a sparse sampling of refractive scintillation cycles can therefore somewhat bias our flux density measurements, and this is reflected by the scatter in Figure \ref{fig:flux_plot}. Nonetheless, since our observations were made in a distributed sub-band setup (i.e. 12$\times$2.56\,MHz spread over the $80$-$220$\,MHz range), our measurements at any given epoch are essentially independent (i.e. 3$\times$12 independent measurements that constrain the low-frequency part of the spectrum in Figure \ref{fig:flux_plot}). The combination of multiple independent measurements, and multiple observations with a time separation much longer than the RISS time scale, effectively works to decrease any possible bias. Indeed, more observational data will help further refine these measurements and thus yield more robust estimates of the spectrum.

Our measured spectral index is consistent with one of the two estimates ($\alpha_{1}$\,=\,${-}2.12$\,$\pm$\,0.04) from \citet{2015Dai}, which is the best fit to their observations made at three Parkes bands. \citet{2015Dai} also quote a much larger value for the spectral index, $\alpha_{2}$\,=\,${-}2.93$\,$\pm$\,0.07, when suitable weights are applied to longer observations. Incidentally, all these estimates are discrepant with the estimate from \citet{2017Murphy}, who report $\alpha$\,=\,${-}1.3$\,$\pm$\,0.1. This discrepancy is likely due to their use of comparatively less reliable measurements (at $\sim 1$\,GHz) and the measurements from continuum imaging data (at lower frequencies).

A closer examination of Parkes data suggests that the spectrum is steeper for most of the main pulse ($-3.0$\,$\textless$\,$\alpha$\,$\textless$\,${-}2.5$), and shallower towards the trailing side of the main pulse ($\textgreater$\,${-}2.5$; see Figure A24, \citealp{2015Dai}).
Given the prominence of precursor emission ($\sim$ 10$-$15\% of the pulse peak, which corresponds to $\sim$ 380\,mJy at 158\,MHz) in the MWA band, we can place an upper limit on its spectral index ($\rm \alpha_{\rm PC}$\,$\textless$\,${-}3.7$), which may indicate a likely spectral steepening at low frequencies. The other plausible cause is possible frequency evolution of the pulse profile, which would require further detailed investigations using the wideband instrumentation now available at the Parkes and Giant Metrewave Radio Telescopes, preferably from observations conducted simultaneously with the MWA.

The recent discoveries of MSPs at low frequencies (\citealp{2017Bassa, 2018Lynch}) re-affirms the general trend of MSPs exhibiting steeper spectra, which is also consistent with our recent work (e.g., \citealp{2018Bhat}).
This is encouraging from the perspective of finding new pulsars using current and next-generation low-frequency telescopes such as the low-frequency SKA.

\section{Conclusions}\label{sec:future_work}

The development of a new software beamformer for MWA pulsar processing has enabled high-quality low-frequency detections of the short-period MSP {\PSR}, a top-priority target for PTA applications. By inverting the polyphase filter-bank operation that was performed prior to the VCS recording, the new beamformer produces high-time resolution ($\sim$\,1\,$\rm \mu s$) voltage time series data that are amenable for phase-coherent de-dispersion. With the resultant high-fidelity detections, we are able to achieve a timing precision of the order $\sim$\,1\,$\rm \mu s$ at the low frequencies of the MWA, which allowed us to measure the pulsar DM with a very high precision, of the order of $(2$-$6)$\,$\times$\,$10^{-6}$\,{\DMunit}. This provides an excellent demonstration of the MWA's capability to monitor subtle DM variations ($\sim 10^{-4}$\,{\DMunit}) reliably for southern pulsars. The DM precision, and the sensitivity to DM changes, can be further improved in future as more observations accrue and the use of frequency-dependent templates becomes possible. Given PSR {\PSR} is fast emerging as one of the high-priority targets for current and future PTAs, DM excess and/or variation at the level of $\sim$\,$10^{-4}$\,{\DMunit} deserves particular importance, as any residual uncorrected DM variations can impact the achievable timing precision.

Our observations also reveal a dual-precursor type feature in the pulse profile, which appears to be far less prominent at higher frequencies from Parkes. The precursor emission appears to have a steeper spectrum than that of the main pulse. The estimated spectral index for pulsar's main pulse emission $\rm \alpha_{MP}$\,=\,${-}2.25$\,$\pm$\,$0.10$, and there is no sign of any turnover within the observed MWA band. This is consistent with the general trend from other recent studies of MSPs and is promising for their low-frequency studies in general, particularly for finding new ones at low frequencies.

\textit{Acknowledgements:} \label{sec:Acknowledgements} \\
We would like to thank an anonymous referee for valuable comments and suggestions that helped improve the clarity and presentation of this paper. This scientific work makes use of the Murchison Radio-astronomy Observatory, operated by CSIRO. We acknowledge the Wajarri Yamatji people as the traditional owners of the Observatory site. Support for the operation of the MWA is provided by the Australian Government (NCRIS), under a contract to Curtin University administered by Astronomy Australia Limited. We acknowledge the Pawsey Supercomputing Centre, which is supported by the Western Australian and Australian Governments. Parts of this research were conducted by the Australian Research Council Centre of Excellence for All-sky Astrophysics (CAASTRO), through project no. CE110001020. DK acknowledges support from the Curtin International Postgraduate Research Scheme. R.M.S. acknowledges support through ARC grant CE170100004. DLK was supported by the NANOGrav project, which receives support from NSF Physics Frontier Center award number 1430284. 

\textit{Facilities}: The MWA and Parkes radio telescope facilities.

\textit{Software}: This work made use of the following software packages: {\tt DSPSR} (\citealp{2011Van&Bailes}) and {\tt PSRCHIVE} (\citealp{2004Hotan, 2012Van})

\bibliographystyle{aasjournal}
\bibliography{2241paper}

\end{document}